\documentclass[
aps,
prb,
amsmath,amssymb,
twocolumn,
superscriptaddress,
showkeys,
showpacs,
nofootinbib,
]{revtex4-1}

\usepackage{bm}
\usepackage{enumitem}
\usepackage{graphicx}
\usepackage[svgnames]{xcolor}
\usepackage{siunitx}
\usepackage[caption=false]{subfig}
\usepackage[colorlinks=true, citecolor=Blue, linkcolor=Blue, urlcolor=Blue]{hyperref}

\usepackage{etoolbox}
\makeatletter
\appto\abstract{
  \let\latexlist\list
  \def\list{\edef\keeprightskip{\the\rightskip}\latexlist}
  \patchcmd\latexlist{\ignorespaces}{\rightskip\keeprightskip\ignorespaces}{}{}
}
\makeatother

\begin{document}

\title{Spontaneously broken translational symmetry at edges of high-temperature superconductors: thermodynamics in magnetic field}

\author{P. Holmvall}
 \email[]{holmvall@chalmers.se}
 \affiliation{Department of Microtechnology and Nanoscience - MC2, Chalmers University of Technology, SE-41296 G{\"o}teborg, Sweden}
 \author{A. B. Vorontsov}
 \email[]{anton.vorontsov@montana.edu}
  \affiliation{Department of Physics, Montana State University, Montana 59717, USA}
\author{M. Fogelstr{\"o}m}
\email[]{mikael.fogelstrom@chalmers.se}
 \affiliation{Department of Microtechnology and Nanoscience - MC2, Chalmers University of Technology, SE-41296 G{\"o}teborg, Sweden}
\author{T. L{\"o}fwander}
\email[]{tomas.lofwander@chalmers.se}
 \affiliation{Department of Microtechnology and Nanoscience - MC2, Chalmers University of Technology, SE-41296 G{\"o}teborg, Sweden}

\date{\today}

\begin{abstract}

We investigate equilibrium properties, including structure of the order parameter, superflow patterns, and thermodynamics of low-temperature surface phases of layered $d_{x^2-y^2}$-wave superconductors in magnetic field.  
At zero external magnetic field, time-reversal symmetry and continuous translational symmetry along the edge are broken spontaneously in a second order phase transition at a temperature $T^*\approx 0.18 T_c$, where $T_c$ is the superconducting transition temperature. At the phase transition there is a jump in the specific heat that scales with the ratio between the edge length $D$ and layer area ${\cal A}$ as $(D\xi_0/{\cal A})\Delta C_d$, where $\Delta C_d$ is the jump in the specific heat at the $d$-wave superconducting transition and $\xi_0$ is the superconducting coherence length. The phase with broken symmetry is characterized by a gauge invariant superfluid momentum $\bm{p}_s$ that forms a non-trivial planar vector field with a chain of sources and sinks along the edges with a period of approximately $12\xi_0$, and saddle point disclinations in the interior.
To find out the relative importance of time-reversal and translational symmetry breaking we apply an external field that breaks time-reversal symmetry explicitly. We find that the phase transition into the state with the non-trivial $\bm{p}_s$ vector field keeps its main signatures, and is still of second order.
In the external field, the saddle point disclinations are pushed towards the edges, and thereby a chain of edge motifs are formed, where each motif contains a source, a sink, and a saddle point. Due to a competing paramagnetic response at the edges, the phase transition temperature $T^*$ is slowly suppressed with increasing magnetic field strength, but the phase with broken symmetry survives into the mixed state.

\end{abstract}

\maketitle

\section{\label{sec:introduction}Introduction}

Superconducting devices are often experimentally realized as thin-film circuits or hybrid structures operating in the mesoscopic regime.\cite{bib:Goltsman_2001,bib:DeFranceschi_2010,bib:Welp_2013,bib:Fornieri_2017} At this length-scale, where the size of circuit elements become comparable with the superconducting coherence length, the nature of the superconducting state may be dictated by various finite-size or surface/interface effects \cite{bib:gustafsson_2013}. This holds true in particular for unconventional superconductors, such as the high-temperature superconductors with an order parameter of $d_{x^2-y^2}$ symmetry that changes sign around the Fermi surface. Scattering at surfaces, or any defect, then leads to substantial pair breaking and formation of Andreev states with energies within the superconducting gap \cite{bib:kashiwaya_tanaka_2000,bib:lofwander_shumeiko_wendin_2001}. Today, the material control of high-temperature superconducting films is sufficiently good that many advanced superconducting devices can work at elevated temperatures \cite{bib:Baghdadi_2015,bib:Xie_2017}. This raises the question how the specific surface physics of $d$-wave superconductors influence devices.

\begin{figure*}[p]
\includegraphics[width=\textwidth]{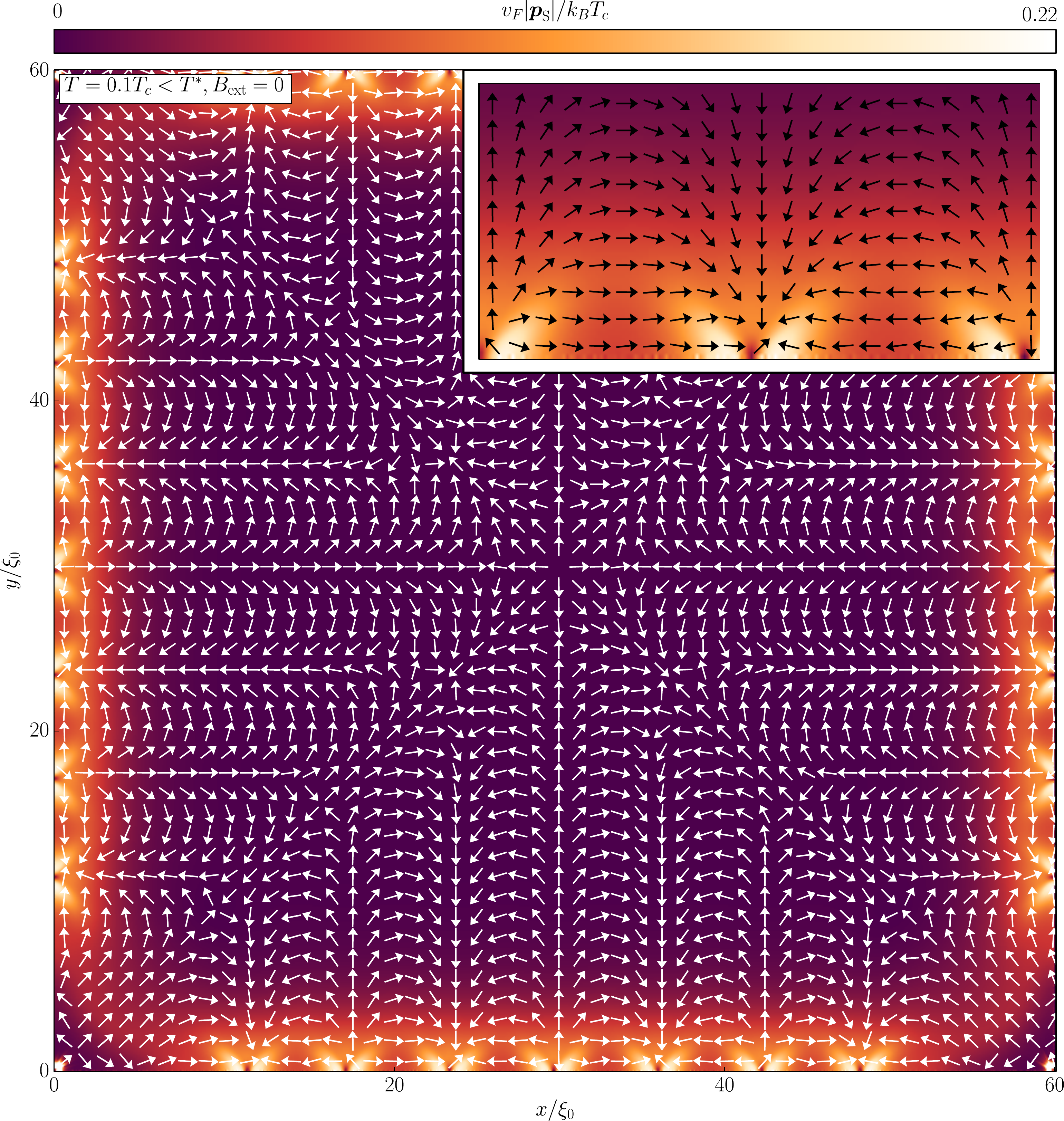}
\caption{For $B_{\text{ext}}=0$, the superfluid momentum $\bm{p}_s$ forms a non-trivial planar vector field with a regular chain of sources and sinks along the edge (each with winding number $n=1/2$), thereby breaking continuous translational symmetry along the edge. Matching saddle point disclinations with winding number $n=-1$ are formed in the interior, along the grain diagonals. Because of the particular grain geometry, four edge sources at the middle of the sides and four corner sources plus an $n=1$ source in the center are matched by four saddle points near the grain center. The temperature is $T=0.1T_c$, which is well below $T^*$. As a consequence, the splay patterns are rather stiff, leading to triangular shapes near the edges. The stiffness is clear from the magnitude variation shown in colorscale and in the inset.}
\label{fig:ps}
\end{figure*}
From a theory point of view, the physics at specular pair-breaking surfaces of $d$-wave superconductors is rich and interesting. The reason is the formation of zero-energy (midgap) Andreev states due to the sign change of the $d$-wave order parameter for quasiparticles scattered at the surface\cite{bib:hu_1994,bib:kashiwaya_tanaka_2000,bib:lofwander_shumeiko_wendin_2001}. In modern terms, there is a flat band of spin-degenerate zero-energy surface states as function of the parallel component of the momentum, $p_{||}$, which is a good quantum number for a specular surface. The large spectral weight of these states exactly at zero energy (i.e. at the Fermi energy), is energetically unfavorable. Different scenarios have been proposed, within which there is a low-temperature instability and a phase transition into a time-reversal symmetry broken phase where the flat band is split to finite energies, thus lowering the free energy of the system.
One scenario is
presence of a subdominant pairing interaction and appearance of another order parameter component $\pi/2$ out of phase with the dominant one 
\cite{bib:Matsumoto_1995a,bib:Matsumoto_1995b,bib:fogelstrom_rainer_sauls_1997,bib:Sigrist_1998}, for instance a subdominant $s$-wave resulting in an order parameter combination $\Delta_d+i\Delta_s$. The phase transition is driven by a split of the flat band of Andreev states to $\pm\Delta_s$. The split Andreev states carry current along the surface, which results in a magnetic field that is screened from the bulk. In a second scenario, exchange interactions drive a ferromagnetic transition at the edge where the flat Andreev band is instead spin split\cite{bib:Honerkamp_2000,bib:Potter_2014}. A third scenario involves spontaneous appearance of supercurrents\cite{bib:higashitani_1997,bib:barash_kalenkov_kurkijarvi_2000,bib:lofwander_shumeiko_wendin_2000} that Doppler shifts the Andreev states and thereby lowers the free energy.
This scenario involves coupling of the electrons to the electromagnetic gauge field $\bm{A}(\bm{R})$,
and was first considered theoretically for a translationally invariant edge.
In this case the transition is a result of the interplay of weakly Doppler shifted surface bound states, decaying away from the surface on the scale of the superconducting coherence length $\xi_0$,  and weak diamagnetic screening currents, decaying on the scale of the penetration depth $\lambda$.
The resulting transition temperature is very low, of order $T^*\sim (\xi_0/\lambda)T_c$, where $T_c$ is the $d$-wave superconducting transition temperature.
Later, the transition temperature was shown to be enhanced in a film geometry\cite{bib:vorontsov_2009,bib:hachiya_aoyama_ikeda_2013,bib:higashitani_miyawaki_2015,bib:miyawaki_higashitani_2015_b,bib:miyawaki_higashitani_2015_a} where two parallel pair breaking edges are separated by a distance of the order of a few coherence lengths. The suppression of the order parameter between the pairbreaking edges can be viewed as an effective Zeeman field that splits the Andreev states and enhances the transition temperature.
The mechanism does not involve subdominant channels or coupling to magnetic field, but depends on film thickness $D$,
and the transition temperature decays rapidly with increasing thickness as $T^*\sim (\xi_0/D) T_c$. In this paper we consider a modified scenario\cite{bib:hakansson_2015,bib:holmvall_2017} where spontaneous supercurrents also break translational symmetry along the edge. This scenario too does not rely on any additional interaction term in the Hamiltonian. Instead, as we will discuss below, it relies on the development of a texture in the gradient of the $d$-wave order parameter phase $\chi$, or more precisely in the gauge invariant superfluid momentum
\begin{equation}
\bm{p}_s(\bm{R})=\frac{\hbar}{2}\nabla\chi(\bm{R})-\frac{e}{c}\bm{A}(\bm{R}),
\label{eq:ps}
\end{equation}
where $\hbar$ is Planck's constant, $e$ the charge of the electron, $c$ the speed of light, and $\bm{A}$ the vector potential. This superfluid momentum spontaneously takes the form of a planar vector field with a chain of sources and sinks along the edge and saddle point disclinations in the grain interior, see Fig.~\ref{fig:ps}. The free energy is lowered by a large split of the flat band of Andreev states  by a Doppler shift $\bm{v}_F\cdot\bm{p}_s$, where $\bm{v}_F$ is the Fermi velocity. This free energy gain is maximized by maximizing the magnitude of $\bm{p}_s$, which is achieved by the peculiar vector field in Fig.~\ref{fig:ps}. The balance of the Doppler shift gain and the energy cost in the disclinations with $\nabla\times\bm{p}_s\neq 0$ and the splay patterns between them leads to a high $T^*\approx 0.18T_c$.  The inhomogeneous vector field induces a chain of loop-currents at the edge circulating clockwise and anti-clockwise. The induced magnetic fluxes of each loop are a fraction of the flux quantum and forms a chain of fluxes with alternating signs along the edge. In this paper we study the thermodynamics of this phase in more depth and investigate the influence of an external magnetic field, explicitly breaking time-reversal symmetry. As we shall see, in the external magnetic field, the phase transition is still of second order and is still characterized by the non-trivial vector field $\bm{p}_s(\bm{R})$ breaking continuous translational symmetry along the edge.

Which of these outlined scenarios wins will ultimately depend on material properties of a specific high-temperature superconducting sample, or material properties of other candidate $d$-wave superconductors, e.g. FeSe. In the third scenario, studied in Ref.~\onlinecite{bib:hakansson_2015,bib:holmvall_2017} and in this paper, the resulting transition temperature is large, $T^*\sim 0.18T_c$. It means that the interaction terms in the Hamiltonian for the other scenarios would have to be sufficiently large in order to compete. It is even possible that one or another scenario wins in different parts of the material's phase diagram\cite{bib:Honerkamp_2000}.

From an experimental point of view, the surface physics of $d$-wave superconductors is complicated by for instance surface roughness, inhomogeneous stoichiometry, and presence of impurities. The formation of a band of Andreev states centered at zero energy is well established by numerous tunnelling experiments, in agreement with the expectation for $d$-wave symmetry of the order parameter, as reviewed in Refs.~\onlinecite{bib:kashiwaya_tanaka_2000,bib:lofwander_shumeiko_wendin_2001}. One consistent experimental result is that the band is typically quite broad, with a width that saturates at low temperature. On the other hand, the establishment of a time-reversal symmetry breaking phase remains under discussion, see for instance Refs.~\onlinecite{bib:kirtley_2006,bib:saadaoui_2011}. Several tunneling experiments on YBCO \cite{bib:covington_1997,bib:dagan_2001,bib:elhalel_2007} show a split of the zero-bias conductance peak, while others do not\cite{bib:neils_2002,bib:deutscher_2005}. Other probes indicating time-reversal symmetry breaking include thermal conductivity\cite{bib:krishana_1997}, Coulomb blockade in nanoscale islands\cite{bib:gustafsson_2013}, and STM tunneling at grain boundaries in FeSe\cite{bib:watashige_2015}.

As we argued in Refs.~\onlinecite{bib:hakansson_2015,bib:holmvall_2017} within the scenario with spontaneous loop currents, the split of the Andreev band might be difficult to resolve in a tunneling experiment because of the 
broken translational symmetry along the edge and associated variations in the superflow field.
This leads to a smearing effect for tunnel contacts with an area larger than the coherence length and an expected wide, largely temperature-independent, peak centered at zero energy. In fact, this would be consistent with most tunneling experiments. With an eye to inspire a new generation of experiments, we present results for the interplay between an external magnetic field,
that induces screening supercurrents,
and the phase transition at $T^*$ into a state with the spontaneous loop currents at the edges. After a brief overview in Section~\ref{sec:model} of the quasiclassical formalism that we use, we will in Section~\ref{sec:results} present results that show this interplay from different points of view. First, we will show the spontaneous currents and induced magnetic fields and relate them to the disclinations in the superfluid momentum; second, we study the magnetic field dependent thermodynamics of the phase transition. Finally in Section~\ref{sec:conclusion}, we summarize our results and provide an outlook.

\section{\label{sec:model}Theoretical model}

\begin{figure}[t]
\includegraphics[width=1.0\linewidth]{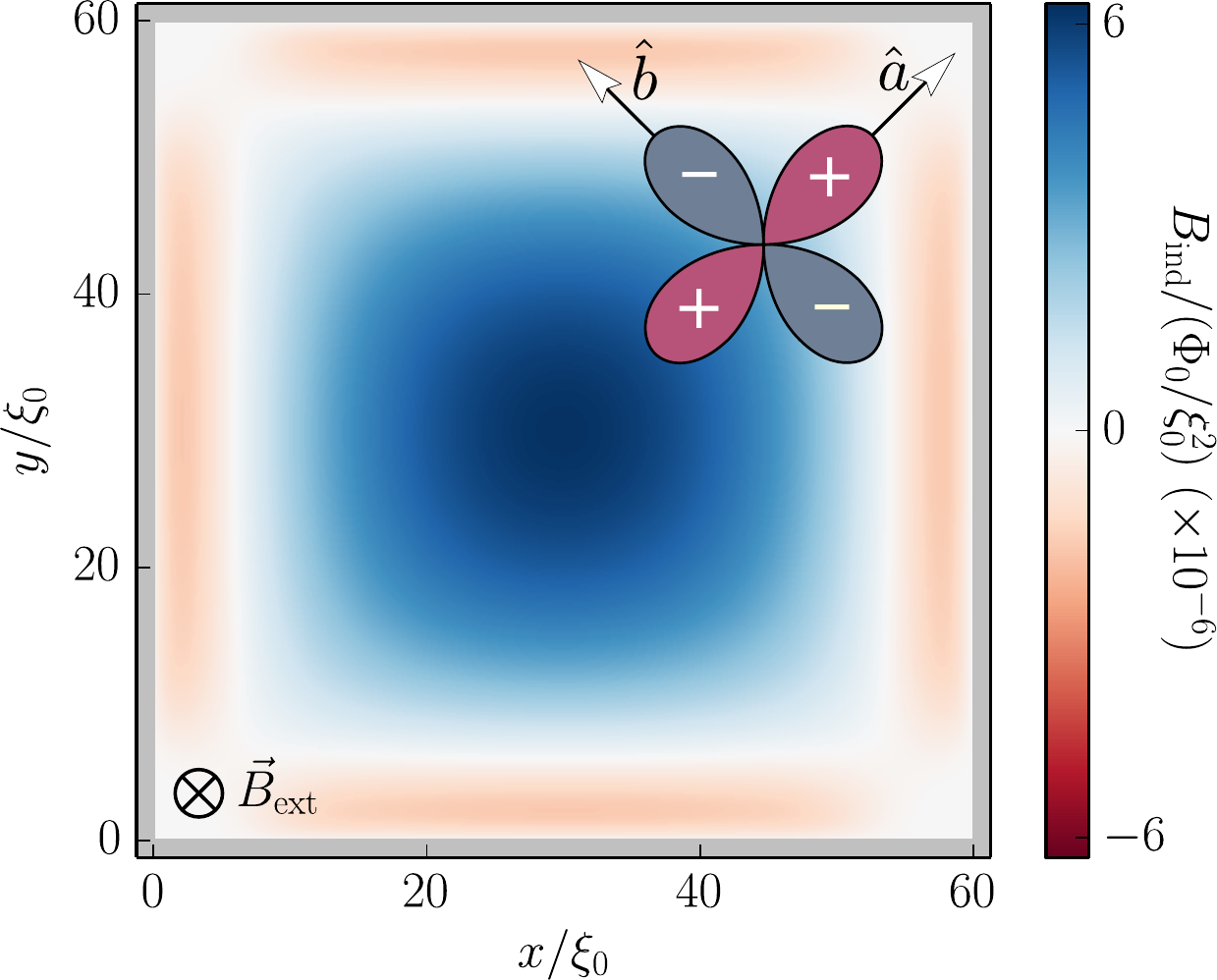}
\caption{\label{fig:geometry} (Color online) The system consists of a $d$-wave superconducting grain exposed to an external magnetic field $\bm{B}_{\text{ext}} = -B_{\text{ext}} \bm{\hat{z}}$. The crystal $ab$-axes are rotated $45^{\circ}$ relative to the grain edges, inducing pair-breaking at the edges of the system. The color scale shows the magnetic field $B_{\text{ind}}$ induced in response to an external field of size $B_{\text{ext}}=\Phi_0/2{\cal A}$ at a temperature $T=0.2T_c$. There is a diamagnetic response carried by the condensate in the interior, and a paramagnetic response carried by mid-gap surface Andreev states at the edges.}
\end{figure}

Our aim is to investigate the ground state of clean mesoscopic $d$-wave superconducting grains in an external magnetic field applied perpendicular to the crystal $ab$-plane, as shown in Fig.~\ref{fig:geometry}. As a typical geometry we consider a square grain with sidelengths $D = 60\xi_0$, where $\xi_0 = \hbar v_F/(2\pi k_BT_c)$ is the zero-temperature superconducting coherence length. Here, $v_F$ is the normal state Fermi velocity, and $k_B$ the Boltzmann constant. The sides of the system are assumed to be misaligned by a $45^{\circ}$ rotation with respect to the crystal $ab$-axes, inducing maximal pair-breaking at the edges.

The external field is directed perpendicular to the $xy$-plane,
\begin{equation}
\label{eq:theory:b}
\bm{B}_{\text{ext}} = -B_{\text{ext}}\bm{\hat{z}} \parallel \bm{\hat{c}}.
\end{equation}
We shall consider rather small external fields, and will use a field scale $B_{g1}=\Phi_0/{\cal A}$, corresponding to one flux quantum threading the grain of area ${\cal A}=D^2=60\xi_0 \times 60\xi_0$. The flux quantum $\Phi_0 = hc/(2|e|)$ is given in Gaussian CGS units. The field $B_{g1}$ is larger than the lower critical field $B_{c1}\propto \Phi_0/\lambda_0^2$, where vortices can enter a macroscopically large superconductor, since the grain side length is smaller than the penetration depth. We assume that $\lambda_0=100\xi_0$, relevant for YBCO).
The upper critical field $B_{c2}\propto \Phi/\xi_0^2$ is much larger than any field we include in this study.
To be precise, we parameterize the field strength as
\begin{equation}
\label{eq:theory:b_units}
B_{\text{ext}} = b B_{g1}, \,\, B_{g1}\equiv \frac{\Phi_0}{\cal A},
\end{equation}
and will consider $b\in[0,1.5]$. 

\subsection{\label{sec:model:quasiclassics}Quasiclassical theory}
We utilize the quasiclassical theory of superconductivity \cite{bib:eilenberger_1968,bib:larkin_ovchinnikov_1969,bib:shelankov_1985}, which is a theory based on a separation of scales \cite{bib:serene_rainer_1983,bib:eschrig_heym_rainer_1994,bib:eschrig_rainer_sauls_1999,bib:sauls_eschrig_2009}.
For instance the atomic scale is assumed small compared with the superconducting coherence length, $\hbar/p_F\ll \xi_0$.
This separation of scales makes it possible to systematically expand all quantities in small parameters such as $\hbar/p_F\xi_0$, $\Delta/\epsilon_F$,  and $k_BT_c/\epsilon_F$, where $\Delta$ is the superconducting order parameter, $p_F$ is the Fermi momentum, and $\epsilon_F$ is the Fermi energy.
In equilibrium, the central object of the theory is the quasiclassical Green's function $\hat{g}(\bm{p}_F,\bm{R};z)$, which is a function of quasiparticle momentum on the Fermi surface $\bm{p}_F$, the quasiparticle center-of-mass coordinate $\bm{R}$, and the quasiparticle energy $z$. The latter is real $z=\epsilon + i0^{+}$ with an infinitesimal imaginary part $i0^{+}$ for the retarded Green's function, or an imaginary Matsubara energy $z = i\epsilon_n = i\pi k_BT(2n+1)$ in the Matsubara technique ($n$ is an integer). To keep the notation compact, the dependence on the parameters $\bm{p}_F$, $\bm{R}$, and $z$ will often not be written out.
The \textit{hat} on $\hat{g}$ denotes Nambu (electron-hole) space
\begin{equation}
\label{eq:theory:green_function}
\hat{g} = 
\left(
\begin{array}{cc}
{g} & {f}\\
-{\tilde{f}} & {\tilde{g}}
\end{array}\right),
\end{equation}
where ${g}$ and ${f}$ are the quasiparticle and pair propagators, respectively.
The tilde operation denotes particle-hole conjugation 
\begin{equation}
\label{eq:theory:tilde}
\tilde{\alpha}(\bm{p}_F,\bm{R};z) = \alpha^{*}(-\bm{p}_F,\bm{R};-z^{*}).
\end{equation}
The quasiclassical Green's function is parameterized in terms of two scalar coherence functions, $\gamma(\bm{p}_F,\bm{R};z)$ and $\tilde \gamma(\bm{p}_F,\bm{R};z)$, as\cite{bib:nagato_1993,bib:schopohl_maki_1995,bib:schopohl_1998,bib:eschrig_sauls_rainer_1999,bib:eschrig_2000,bib:vorontsov_sauls_2003,bib:eschrig_2009}
\begin{equation}
\label{eq:theory:g_parametrization}
\hat{g} = -\frac{i\pi}{1+\gamma\tilde\gamma}
\left(
\begin{array}{cc}
{1}-{\gamma}{\tilde{\gamma}} & 2{\gamma}\\
2{\tilde{\gamma}} & -{1}+{\gamma}{\tilde \gamma}
\end{array}\right).
\end{equation}
Note that with this parameterization, the Green's function is automatically normalized to $\hat g^2 = -\pi^2$.
The coherence functions obey two Riccati equations
\begin{eqnarray}
\label{eq:theory:riccati_gamma}
(i\hbar\bm{v}_F\cdot\bm{\nabla}+2z+2\frac{e}{c}\bm{v}_F\cdot\bm{A}){\gamma} & = & -{\tilde{\Delta}}{\gamma^2}-{\Delta},\\
\label{eq:theory:riccati_gammatilde}
(i\hbar\bm{v}_F\cdot\bm{\nabla}-2z-2\frac{e}{c}\bm{v}_F\cdot\bm{A})\tilde{\gamma} & = & -{\Delta}{\tilde{\gamma}^2}-{\tilde{\Delta}},
\end{eqnarray}
where $\bm{A}$ is the vector potential.
These first order nonlinear differential equations are solved by integration along straight (ballistic) quasiparticle trajectories. Quantum coherence is retained along these trajectories, but not between neighboring trajectories. A clean superconducting grain in vacuum is assumed by imposing the boundary condition of perfect specular reflection of quasiparticles along the edges of the system.

The superconducting order parameter is assumed to have pure $d$-wave symmetry
\begin{equation}
\label{eq:theory:order_parameter}
\Delta(\bm{p}_F,\bm{R}) = \Delta_d(\bm{R})\eta_d(\theta),
\end{equation}
where $\theta$ is the angle between the Fermi momentum $\bm{p}_F$ and the crystal ${\bf\hat a}$-axis, and $\eta_d(\theta)$ is the $d$-wave basis function
\begin{equation}
\label{eq:theory:basis_function}
\eta_{d}(\theta) = \sqrt{2}\cos(2\theta),
\end{equation}
fulfilling the normalization condition
\begin{equation}
\label{eq:theory:basis_function:normalization}
\int\frac{d\theta}{2\pi}\left|\eta_{d}(\theta)\right|^2 = 1.
\end{equation}
The order parameter amplitude satisfies the gap equation
\begin{equation}
\label{eq:theory:gap_equation}
\Delta_{d}(\bm{R}) = \lambda_d N_F k_BT\sum_{|\epsilon_{n}|\leq\Omega_c}
\int \frac{d\theta}{2\pi}\eta^{*}_{d}(\theta)f(\bm{p}_{F},\bm{R};\epsilon_{n}),
\end{equation}
where $\lambda_d$ is the pairing interaction, $N_F$ is the density of states at the Fermi level in the normal state, and $\Omega_c$ is a cutoff energy. The pairing interaction and the cutoff energy are eliminated in favor of the superconducting transition temperature $T_c$ (see for example Ref.~\onlinecite{bib:grein_lofwander_eschrig_2013}) as
\begin{equation}
\label{eq:theory:pairing_interaction}
\frac{1}{\lambda_d N_F} = \ln\frac{T}{T_c} + \sum_{n\geq0}\frac{1}{n+\frac{1}{2}}.
\end{equation}
The above equations are solved self-consistently with respect to $\gamma$, $\tilde\gamma$, and $\Delta_d$. As an initial condition, a homogeneous superconductor is assumed at the start of the trajectories. Along the trajectory and after a few self-consistent iterations, the information of the initial condition is lost \cite{bib:nagai_tanaka_hayashi_2012}.

We choose an electromagnetic gauge where the vector potential has the form
\begin{equation}
\label{eq:theory:vector_potential_external}
\bm{A}_{\text{ext}}(\bm{R}) = \frac{1}{2}\bm{B}_{\text{ext}}\times\bm{R}.
\end{equation}
The total vector potential $\bm{A}(\bm{R})$, that enters Eqs.~(\ref{eq:theory:riccati_gamma})-(\ref{eq:theory:riccati_gammatilde}), is given by $\bm{A}_{\text{ext}}(\bm{R})$ and the field $\bm{A}_{\text{ind}}(\bm{R})$ induced by the currents $\bm{j}(\bm{R})$ in the superconductor [Eq.(\ref{eq:theory:current_density}) below]:  
\begin{equation}
\label{eq:theory:vector_potential_internal}
\bm{A}(\bm{R}) = \bm{A}_{\text{ext}}(\bm{R})+\bm{A}_{\text{ind}}(\bm{R}).
\end{equation}
The vector potential $\bm{A}_{\text{ind}}(\bm{R})$ should be solved from Amp{\`e}re's circuit law
\begin{equation}
\label{eq:theory:ampere}
\bm{\nabla}\times\bm{\nabla}\times\bm{A}_{\text{ind}}(\bm{R}) = \frac{4\pi}{c}\bm{j}(\bm{R}).
\end{equation}
To take the full electrodynamics into account, $A_{\text{ind}}(\bm{R})$ also needs to be computed self-consistently in each iteration.
However, the strength of the electrodynamic back-coupling scales as $\kappa^{-2}$, where $\kappa \equiv \lambda_0/\xi_0$ is the dimensionless Ginzburg-Landau parameter. The electrodynamic back-coupling can therefore safely be neglected for type II superconductors (typically $\kappa^{-1} \approx 10^{-2}$ for the cuprates).

The induced magnetic flux density is computed as
\begin{equation}
\bm{B}_{\text{ind}} = \bm{\nabla}\times\bm{A}_{\text{ind}}.
\end{equation}
We shall neglect the problem of the field distribution around the superconductor and focus on the field induced at the $ab$-plane where we have simply $\bm{B}_{\text{ind}}=B_{\text{ind}}\bm{\hat{z}}$.

\subsection{\label{sec:model:ps}{Gauge transformation}}

Once the Green's function and the order parameter have been determined self-consistently, we can perform a gauge transformation in order to make the order parameter a real quantity and in the process extract the superfluid momentum $\bm{p}_s$. This can be illustrated by transforming the Riccati equation in Eq.~(\ref{eq:theory:riccati_gamma}). To begin with, the self-consistently obtained order parameter is complex, i.e.
\begin{equation}
\Delta(\bm{p}_F,\bm{R}) = |\Delta_d(\bm{R})|\eta_d(\theta)\,e^{i\chi(\bm{R})}.
\end{equation}
We make the ansatz
\begin{equation}
\gamma(\bm{p}_F,\bm{R};z) = \gamma_0(\bm{p}_F,\bm{R};z)\,e^{i\chi(\bm{R})},
\end{equation}
and put that into the Riccati equation. We obtain
\begin{equation}
\left[ i\hbar\bm{v}_F\cdot\bm{\nabla} + 2(z-\bm{v}_F\cdot\bm{p}_s) \right] \gamma_0 = -|\Delta_d|\eta_d(\gamma_0^2+1),
\end{equation}
where $\bm{p}_s$ is defined in Eq.~(\ref{eq:ps}).

\subsection{\label{sec:model:observables}Observables}
The current density is computed within the Matsubara technique through the formula
\begin{equation}
\label{eq:theory:current_density}
\bm{j}(\bm{R}) = 2\pi eN_Fk_BT\sum_{\epsilon_n}
\int \frac{d\theta}{2\pi} \bm{v}_F g(\bm{p}_F,\bm{R};\epsilon_n).
\end{equation}
In the results section we shall show this current density in units of the depairing current
\begin{equation}
\label{eq:results:depairing_current}
j_d \equiv 4\pi |e| k_BT_c N_F v_F.
\end{equation}


The free-energy difference between the superconducting and the normal states is calculated with the Eilenberger free-energy functional \cite{bib:eilenberger_1968}
\begin{widetext}
\begin{eqnarray}
\label{eq:theory:eilenberger_free_energy}
\Omega_S(B,T)-\Omega_N(B,T) &=& \int d\bm{R}\left\{ \frac{\bm{B}_{\text{ind}}(\bm{R})^2}{8\pi} + \left|\Delta(\bm{R})\right|^2N_F\ln\frac{T}{T_c} + 2\pi N_Fk_BT\sum_{\epsilon_n>0}\left[\frac{\left|\Delta(\bm{R})\right|^2}{\epsilon_n} + i\mathcal{I}(\bm{R};\epsilon_n)\right]\right\},\\
\mathcal{I}(\bm{R},\epsilon_n) &=& \int\frac{d\theta}{2\pi}
\left[ \tilde\Delta(\bm{p}_F,\bm{R})\gamma(\bm{p}_F,\bm{R};\epsilon_n) - \Delta \tilde\gamma(\bm{p}_F,\bm{R};\epsilon_n) \right].
\end{eqnarray}
\end{widetext}
We have verified that this form of the free energy give the same results as the Luttinger-Ward functional \cite{bib:serene_rainer_1983,bib:vorontsov_sauls_2003,bib:hakansson_2015}. The entropy and specific heat capacity are obtained from the thermodynamic definitions
\begin{eqnarray}
\label{eq:theory:entropy}
S & = & -\frac{\partial \Omega}{\partial T},\\
\label{eq:theory:specific_heat_capacity}
C & = & T\frac{\partial S}{\partial T} = -T\frac{\partial^2 \Omega}{\partial T^2}.
\end{eqnarray}

\section{\label{sec:results}Results}

\begin{table}[b]
\begin{tabular*}{\columnwidth}{p{0.15\columnwidth}p{0.4\columnwidth}p{0.4\columnwidth}}
\hline
\hline        Set              &  Temperature                      & External magnetic field\\
\hline	\textbf{(I)}    & $T=0.182T_c>T^*$             & $B_{\text{ext}} = 0.5B_{g1}$\\
\hline	\textbf{(II)}   & $T=0.176T_c\gtrsim T^*$   & $B_{\text{ext}} = 0.5B_{g1}$\\
\hline	\textbf{(III)}  & $T=0.17T_c< T^*$               & $B_{\text{ext}} = 0.5B_{g1}$\\
\hline	\textbf{(IV)}  & $T=0.17T_c< T^*$              & $B_{\text{ext}} = 0$\\
\hline
\end{tabular*}
\caption{Sets of parameters used for presenting results. The field scale $B_{g1}=\Phi_0/{\cal A}$ corresponds to an external magnetic flux through the grain area exactly equal to one flux quantum.}
\label{table:parameters}
\end{table}

\begin{figure*}[p]
\includegraphics[width=\textwidth,height=\textheight,keepaspectratio]{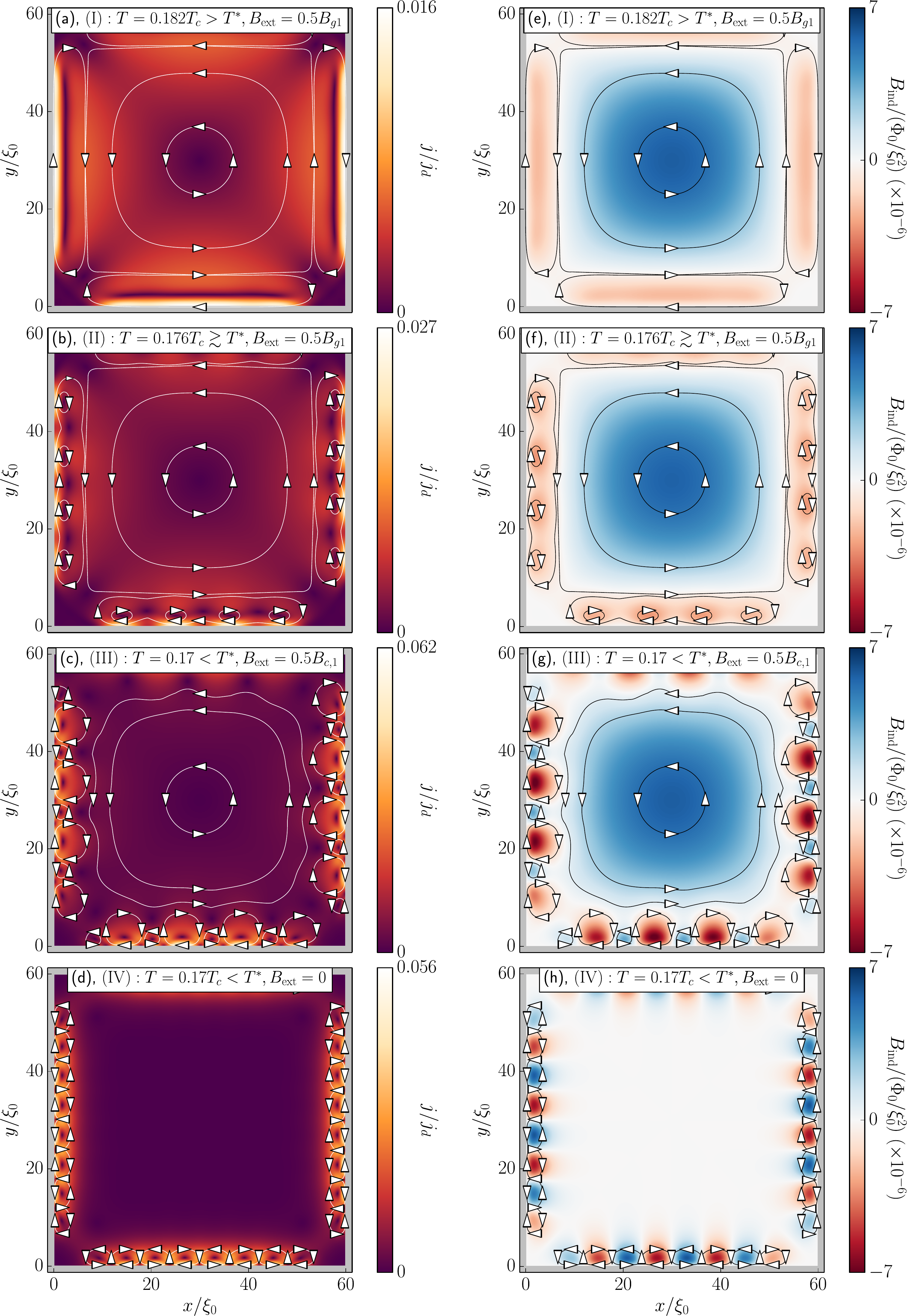}
\caption{\label{fig:fields} (Color online) (a)--(d) Total current magnitude and (e)--(h) induced magnetic flux density for different temperatures and external fields (see annotations). Lines and arrows have been added to illustrate the flow of the currents. }
\end{figure*}

\begin{figure*}[t]
\includegraphics[width=\textwidth]{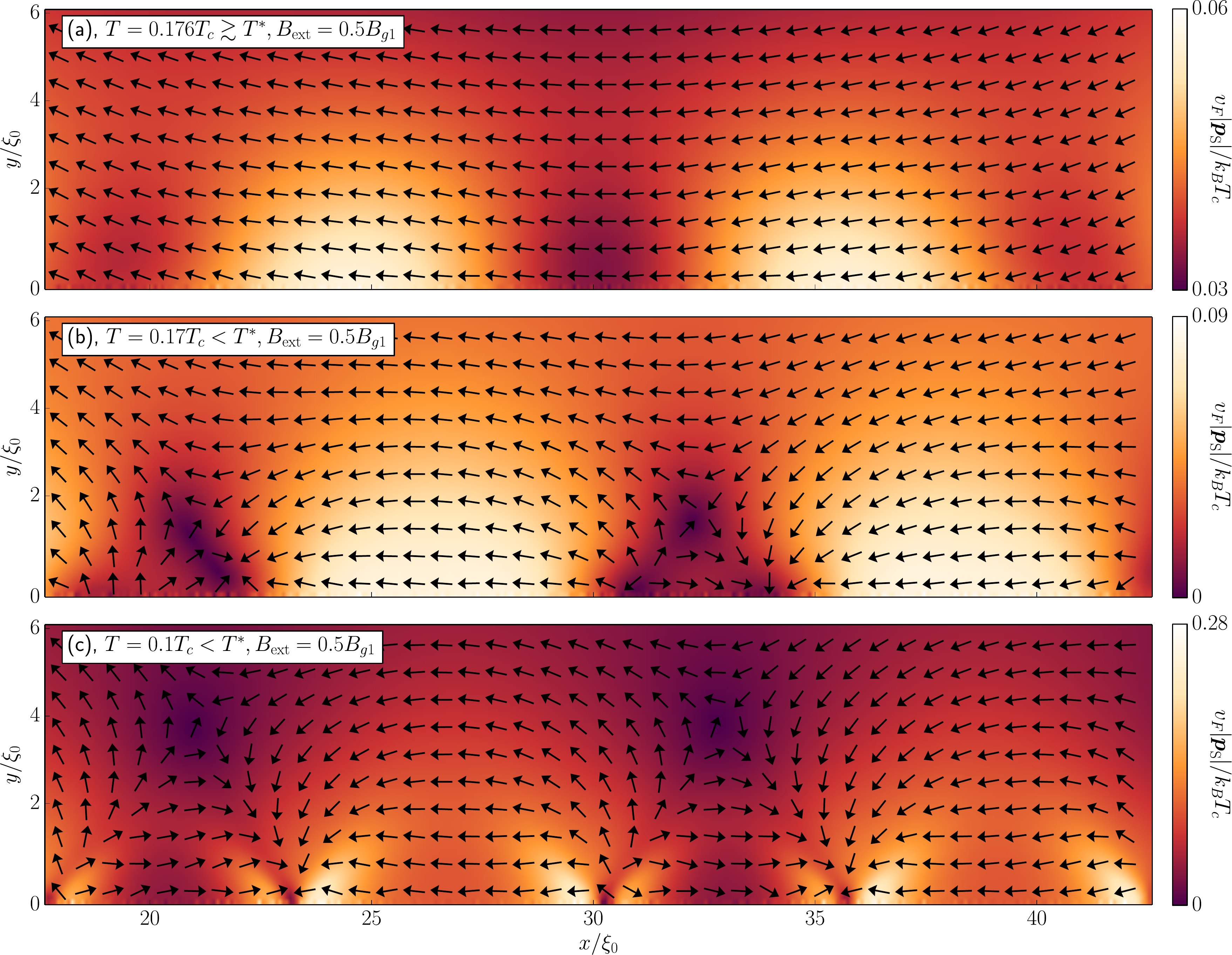}
\caption{The superfluid momentum induced in an external magnetic field of $B_{\text{ext}}=0.5B_{g1}$ for lowering temperatures from top to bottom. At the phase transition, source-sink-sadle-point motifs appear and separate along the edge breaking translational invariance along the edge coordinate $x$. At the same time the magnitude $|\bm{p}_s|$ grows large. Note the different color scales are used in the subfigures in order to enhance visibility.}
\label{fig:ps2}
\end{figure*}
In Fig.~\ref{fig:fields} we show the influence of a rather weak external magnetic field, $B=0.5B_{g1}$, applied to the $d$-wave superconducting grain with pair breaking edges for varying temperature near the phase transition temperature $T^*$. The left and right columns show the currents and the magnetic field densities, respectively, induced in response to the applied field. To be concrete we discuss a few selected sets of model parameters, as listed in Table~\ref{table:parameters}. First, for $T>T^*$ (parameter set I), the expected diamagnetic response of the condensate in the inner part of the grain is present, see Fig.~\ref{fig:fields}(a) and Fig.~\ref{fig:fields}(e). On the other hand, midgap quasiparticle Andreev surface states respond paramagnetically. This situation is well established theoretically and experimentally through measurements of the competition between the diamagnetic and paramagnetic responses seen as a low-temperature up-turn in the penetration depth \cite{bib:Walter_1998}. Upon lowering the temperature to  $T\gtrsim T^*$ (parameter set II), see Fig.~\ref{fig:fields}(b) and Fig.~\ref{fig:fields}(f), the paramagnetic response at the edge becomes locally suppressed and enhanced, forming a sequence of local minima and maxima in the induced currents and fields. The bulk response is on the other hand relatively unaffected. Finally, as $T<T^*$ (parameter set III), see Fig.~\ref{fig:fields}(c) and Fig.~\ref{fig:fields}(g), the regions of minum current turns into regions with reversed currents. The resulting loop currents with clock-wise and anti-clockwise circulations induces magnetic fluxes along the surface with opposite signs between neighboring fluxes. The situation for $T<T^*$ in an external magnetic field can be compared with the one in zero magnetic field displayed in Fig.~\ref{fig:fields}(d) and Fig.~\ref{fig:fields}(h), and also discussed before in Ref.~\onlinecite{bib:hakansson_2015}. In the presence of the magnetic field, there is an imbalance between positive and negative fluxes, while in zero external magnetic field, the total induced flux integrated over the grain area is zero.

Let us quantify the symmetry broken phase in a magnetic field by plotting the superfluid momentum defined in Eq.~(\ref{eq:ps}), see Fig.~\ref{fig:ps2}.
For $T\gtrsim T^*$ (parameter set II), the amplitude of ${\bf p}_s$ varies along the edge (coordinate $x$), see Fig.~\ref{fig:ps2}(a), reflecting the varying paramagnetic response in Fig.~\ref{fig:fields}(b) and Fig.~\ref{fig:fields}(f). For $T<T^*$ (parameter set III), sources and sinks have appeared pairwise together with a saddle point disclination, see Fig.~\ref{fig:ps2}(b). The left disclinations in the figure are not well developed because of the proximity to the corner. Finally, in Fig.~\ref{fig:ps2}(c), we show the vector field at a lower temperature when the chain of sources, sinks, and saddle points are well established and the magnitude of ${\bf p}_s$ is large, much larger than in the interior part of the grain still experiencing diamagnetism. In a magnetic field, the vector field far from the surface has a preferred direction reflecting the diamagnetic response of the interior grain. This shifts the sources and sinks along the surface, as compared with the regular chain for zero field in Fig.~\ref{fig:ps}, and moves the saddle points to the surface region.

\begin{figure*}[t]
\includegraphics[width=\linewidth]{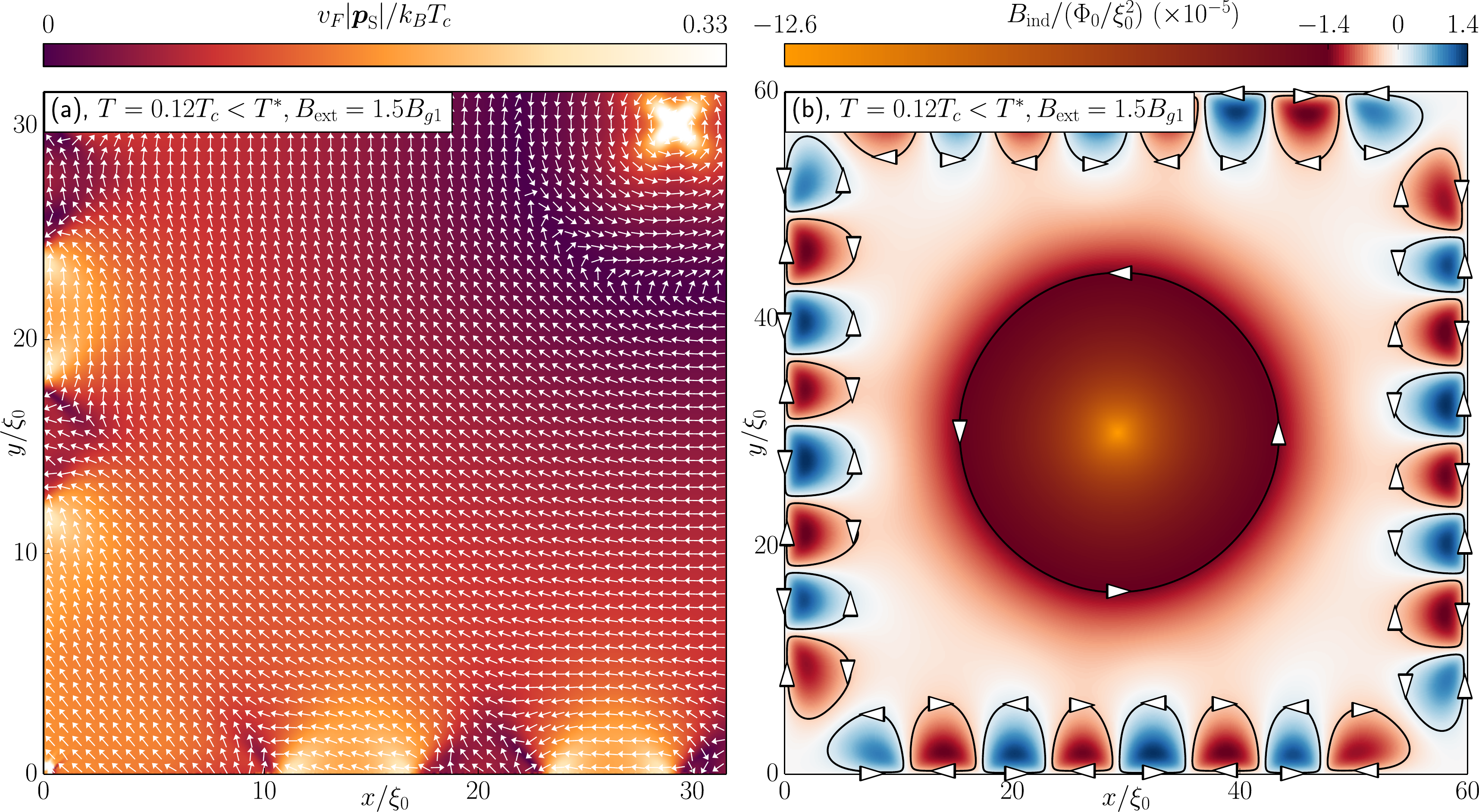}
\caption{\label{fig:vortex_fields} (Color online) (a) Superfluid momentum $\bm{p}_s(\bm{R})$ and (b) the induced field $B_{\text{ind}}(\bm{R})$ at $T=0.1T_c<T^{*}$ and $B_{\text{ext}} = 1.5B_{g1}$, with an Abrikosov vortex stabilized in the center of the grain. The penetration depth is larger than the size of the grain, which means that the Abrikosov vortex is overlapping with the induced fluxes at the pair-breaking edges. The colormap for the flux density has been chosen to show that the edge fluxes (red to blue colors) are similar to those in Fig.~\ref{fig:fields}.}
\end{figure*}

From a vector field perspective,\cite{bib:Mermin_1979,bib:Effenberger_2011} the edge sources and sinks each have a Poincar{\' e} index (winding number) of $n=1/2$. It is not $n=1$ because they lie exactly on the edge. On the other hand, the saddle point has index $n=-1$. Thus, a motif with one edge source, one edge sink, and one saddle point sum up to zero and annihilate at $T^*$. In the same fashion, increasing the magnetic field strength, the motif gets smaller as the disclinations are forced towards each other to match the superflow in the bulk. However, the magnitude of $\bm{p}_s$ near the surface due to Meissner screening of the bulk is not large enough to force an annihilation of the motifs. The broken symmetry phase therefore survives the application of an external magnetic field within the whole Meissner state, $b\in[0,1]$.

In Fig.~\ref{fig:vortex_fields} we show the superfluid momentum for a higher field $B_{\text{ext}}=1.5B_{g1}$, in the mixed state where an Abrikosov vortex resides in the grain center. We find that also in this case the phase with edge loop currents survives. For higher fields, more vortices enter the grain (not shown), still keeping the edge phase intact. However, the exact configuration of Abrikosov vortices becomes geometry dependent and the free energy landscape is very flat. Therefore, the full investigation of the geometry dependent phase diagram for very large fields is beyond the scope of this paper.

\begin{figure}[t]
\includegraphics[width=1.0\linewidth]{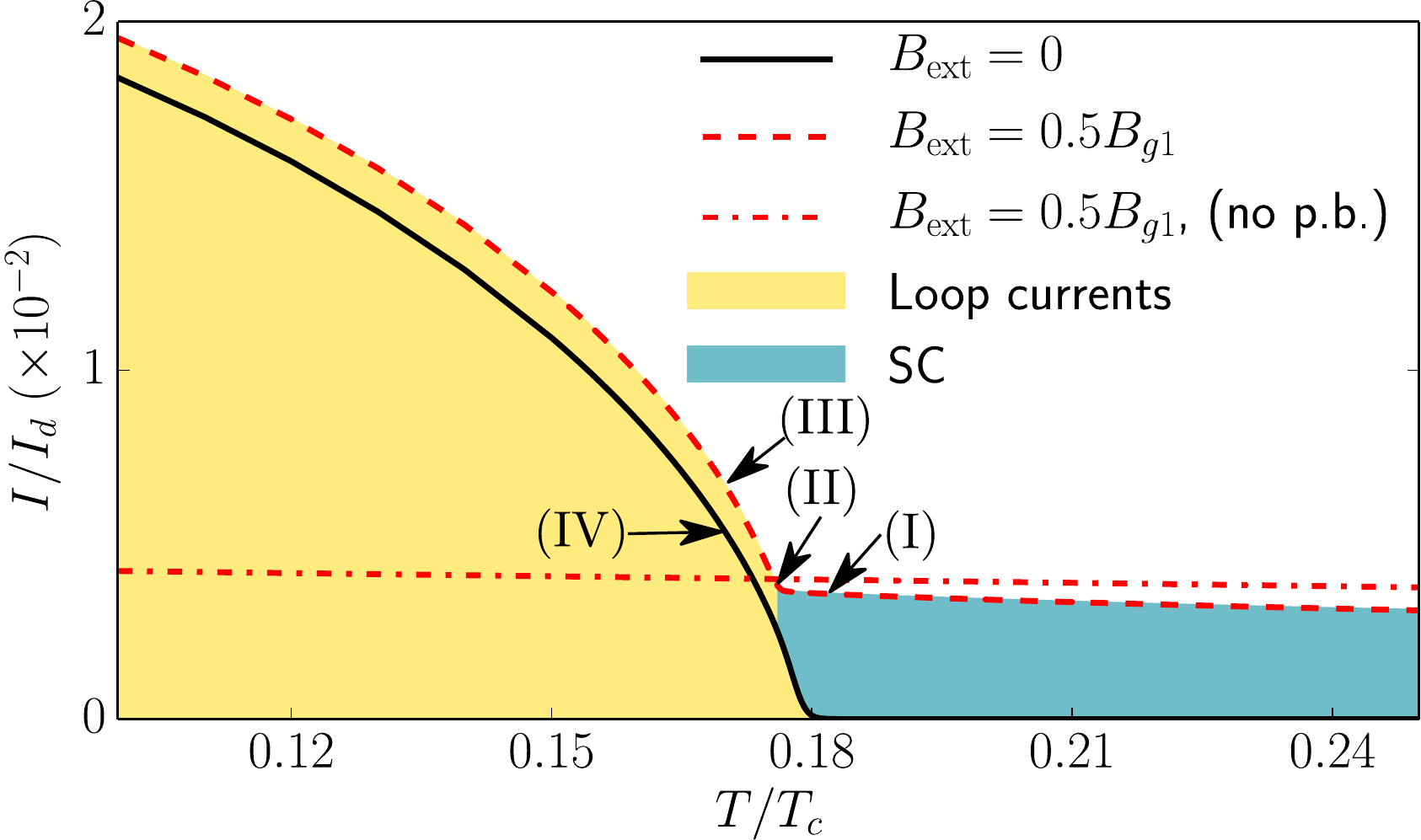}
\caption{\label{fig:int_j} (Color online) Area-averaged current magnitude defined in Eq.~(\ref{eq:j_average}), versus temperature, without any external magnetic field (solid line), with an external magnetic field of magnitude $B_{\text{ext}}=0.5B_{g1}$ (dashed line), and for a system without pair-breaking edges at $B_{\text{ext}}=0.5B_{g1}$ (dash-dotted line). The latter has only a diamagnetic response. Letters (I)--(IV) indicate the parameter values corresponding to the fields in Fig.~\ref{fig:fields}, see Table~\ref{table:parameters}.}
\end{figure}

Let us investigate further how the currents and magnetic fields are induced at $T^*$. As we have seen, the paramagnetic response and the spontaneously appearing edge loop currents compete, as they both lead to shifts of midgap Andreev states. As the temperature is lowered, the strength of the paramagnetic response increases slowly and linearly, while the strength of the loop currents increases highly non-linearly. This is illustrated in Fig.~\ref{fig:int_j}, by plotting the area-averaged current magnitude
\begin{equation}
\overline{j} = \frac{1}{\cal A} \int d^2R\, |\bm{j}(\bm{R})|,
\label{eq:j_average}
\end{equation}
as a function of temperature for the cases when $B_{\text{ext}}=0$ (solid line), $B_{\text{ext}}=0.5B_{g1}$ (dashed line), and for comparison also for a system without pair-breaking edges having only a diamagnetic response at $B_{\text{ext}}=0.5B_{g1}$  (dash-dotted line). The paramagnetic response is fully suppressed at low temperatures $T < T^{*}$. Such a sudden disappearance of the paramagnetic response at a temperature $T^{*}$ should be experimentally measurable, for example in the penetration depth or by using nano-squids \cite{bib:vasyukov_2013,bib:pelliccione_2016}. 

\begin{figure}[t]
\includegraphics[width=1.0\linewidth]{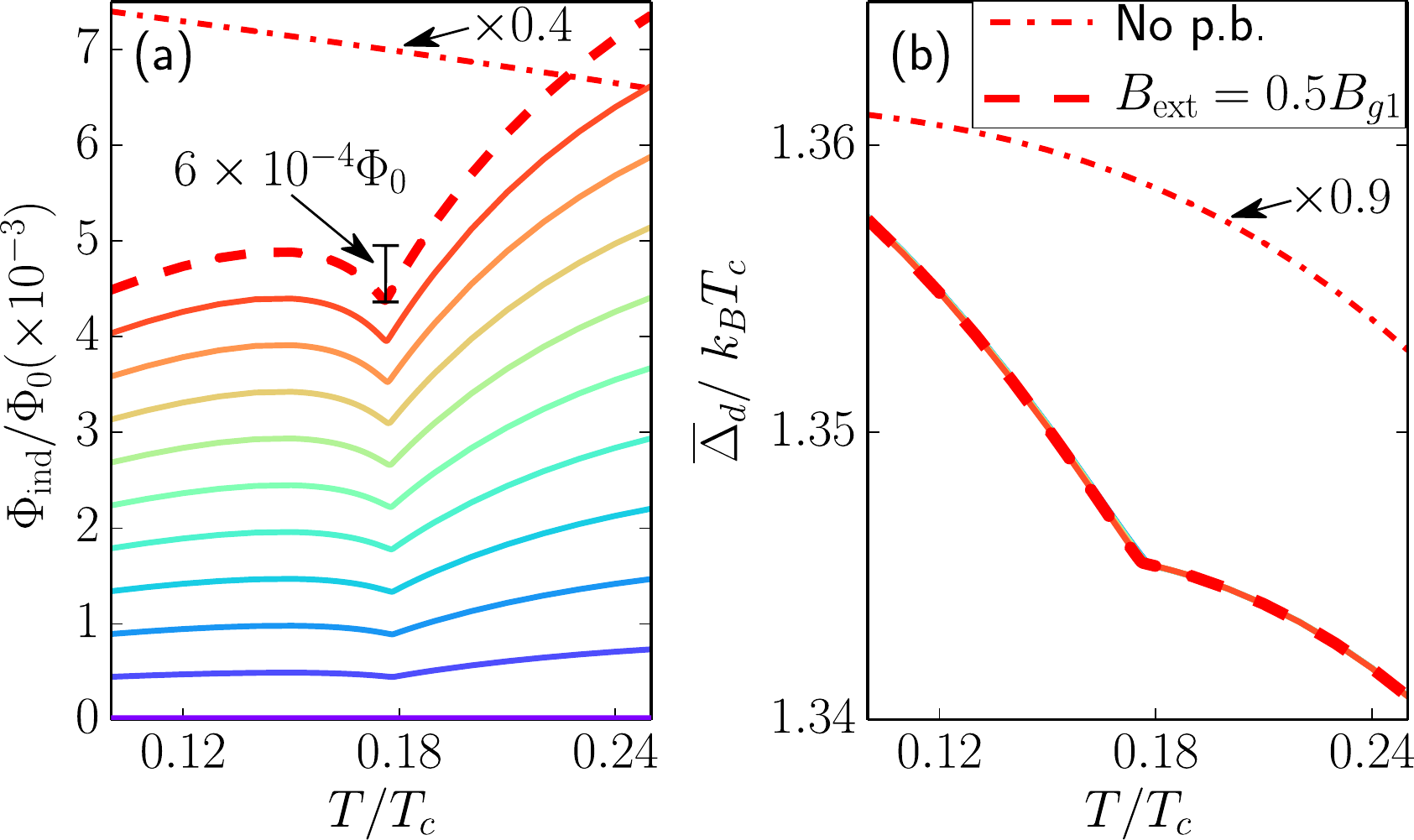}
\caption{\label{fig:int_B} (Color online) (a) Temperature dependence of the induced magnetic flux defined in Eq.~(\ref{eq:Phi_induced}). The solid lines indicate, from bottom to top (colors purple to red), the external field magnitude from $B_{\text{ext}}=0$ to $B_{\text{ext}}=0.5B_{g1}$ in steps of $0.05B_{g1}$. The line corresponding to zero field lies exactly at zero since there is an equal amount of positive and negative fluxes induced in this case, see Fig.~\ref{fig:fields}. Panel (b) shows the area-averaged order parameter magnitude defined in Eq.~(\ref{eq:Delta_average}) versus temperature. Results are also shown for a system without pair-breaking edges (dash-dotted line) at $B_{\text{ext}}=0.5B_{g1}$, but scaled with a factor $0.4$ and $0.9$ in (a) and (b), respectively.}
\end{figure}

We show in Fig.~\ref{fig:int_B}(a) the total induced magnetic flux through the grain
\begin{equation}
\Phi_{\text{ind}} = \int d^2R\, B_{\text{ind}}(\bm{R}),
\label{eq:Phi_induced}
\end{equation}
and in Fig.~\ref{fig:int_B}(b) the area-averaged order parameter magnitude
\begin{equation}
\overline{\Delta}_d =  \frac{1}{\cal A} \int d^2R\, |\Delta_d(\bm{R})|,
\label{eq:Delta_average}
\end{equation}
both as functions of temperature for different values of $B_{\text{ext}}$. The figures also show results for a $d$-wave grain without pair-breaking edges at $B_{\text{ext}}=0.5B_{g1}$ (dash-dotted line). For better visibility, the latter results have been scaled by a factor $0.4$ and $0.9$ in (a) and (b), respectively. Two different trends are distinguishable in the observables for $T<T^*$ and $T>T^*$, separated by a ``kink''. The induced magnetic flux through the grain area decreases as $T$ decreases down to $T^*$ due to the increasing paramagnetic response that competes with the diamagnetic one. At $T^*$, the inhomogeneous edge state appear and starts competing with the paramagnetic response. Thus, the total magnetic flux increases again. At the same time the order parameter is partially healed.

\subsection{\label{sec:results:free_energy}Phase transition and thermodynamics}

\begin{figure*}[t!]
\includegraphics[width=1.0\linewidth]{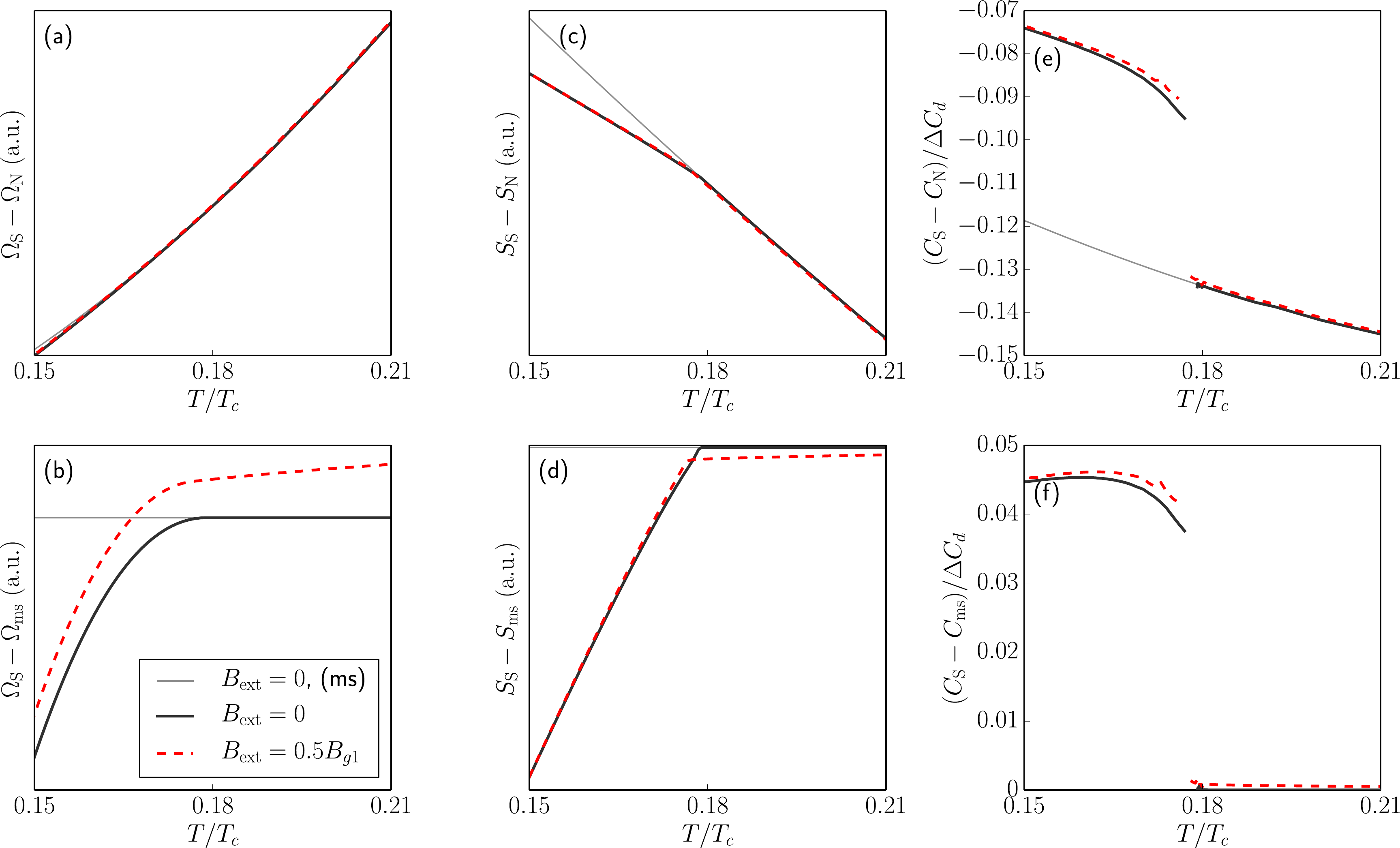}
\caption{\label{fig:free_E} (Color online) (a)--(b) free energy, (c)--(d) entropy, and (e)--(f) specific heat capacity, versus temperature. The lines correspond to a system with purely real order parameter without edge currents (black fine line), a system with spontaneous edge currents in zero magnetic field (black solid line), and in a finite external field $B=0.5B_{g1}$ (red dashed line). In the lower panels (b), (d), and (f), the quantities have been subtracted by the corresponding values of the system with a purely real order parameter, the meta-stable (ms) state. The heat capacity is normalized by the heat capacity jump in the normal-superconducting phase transition for a bulk $d$-wave system, denoted $\Delta C_d$.}
\end{figure*}

The sudden changes with a discontinuity in the derivative as function of temperature of the total induced current, the magnetic flux, as well as the order parameter (Fig.~\ref{fig:int_j}-Fig.~\ref{fig:int_B}) indicate that there is a phase transition occurring at the temperature $T^*$. In zero external magnetic field, we showed in Ref.~\onlinecite{bib:hakansson_2015} that there is a second order phase transition at $T^*$, where both time-reversal symmetry and continuous translational symmetry along the edge are spontaneously broken. Let us now investigate the thermodynamics in an external magnetic field already explicitly breaking time-reversal symmetry.

In Fig.~\ref{fig:free_E}(a) we plot the free energy difference between the superconducting and normal states $\Omega_S-\Omega_N$, defined in Eq.~(\ref{eq:theory:eilenberger_free_energy}), for external field $B=0.5B_{g1}$ (red dashed line) and for zero field (solid black line). For comparison, we show the free energy difference for a purely real order parameter in zero field (black fine line), i.e. without the symmetry breaking edge loop currents. For $T<T^*$, this solution is not the global minimum of the free energy, and we therefore refer to it as a \textit{meta-stable} state. To enhance the visibility of the differences in free energy between the possible solutions, we show in Fig.~\ref{fig:free_E}(b) the free energy difference with respect to the metastable state, i.e. $\Omega_S-\Omega_{ms}$. The small slope in the red dashed line at $T>T^{*}$ in Fig.~\ref{fig:free_E} (b) is caused by the shift of mid-gap Andreev states due to the paramagnetic response, which increases as $T$ decreases. The phase transition temperature $T^*$ for the second order phase transition can be identified with the "knee" in the entropy difference defined in Eq.~(\ref{eq:theory:entropy}), see Fig.~\ref{fig:free_E}(c) and Fig.~\ref{fig:free_E}(d). Since time-reversal symmetry is already explicitly broken by the external magnetic field, the phase transition signals breaking of continuous translational symmetry and establishment of the vector field $\bm{p}_s$ with the chain of disclinations along the edge, as shown in Fig.~\ref{fig:ps2}.

\begin{figure}[t!]
\includegraphics[width=1.0\linewidth]{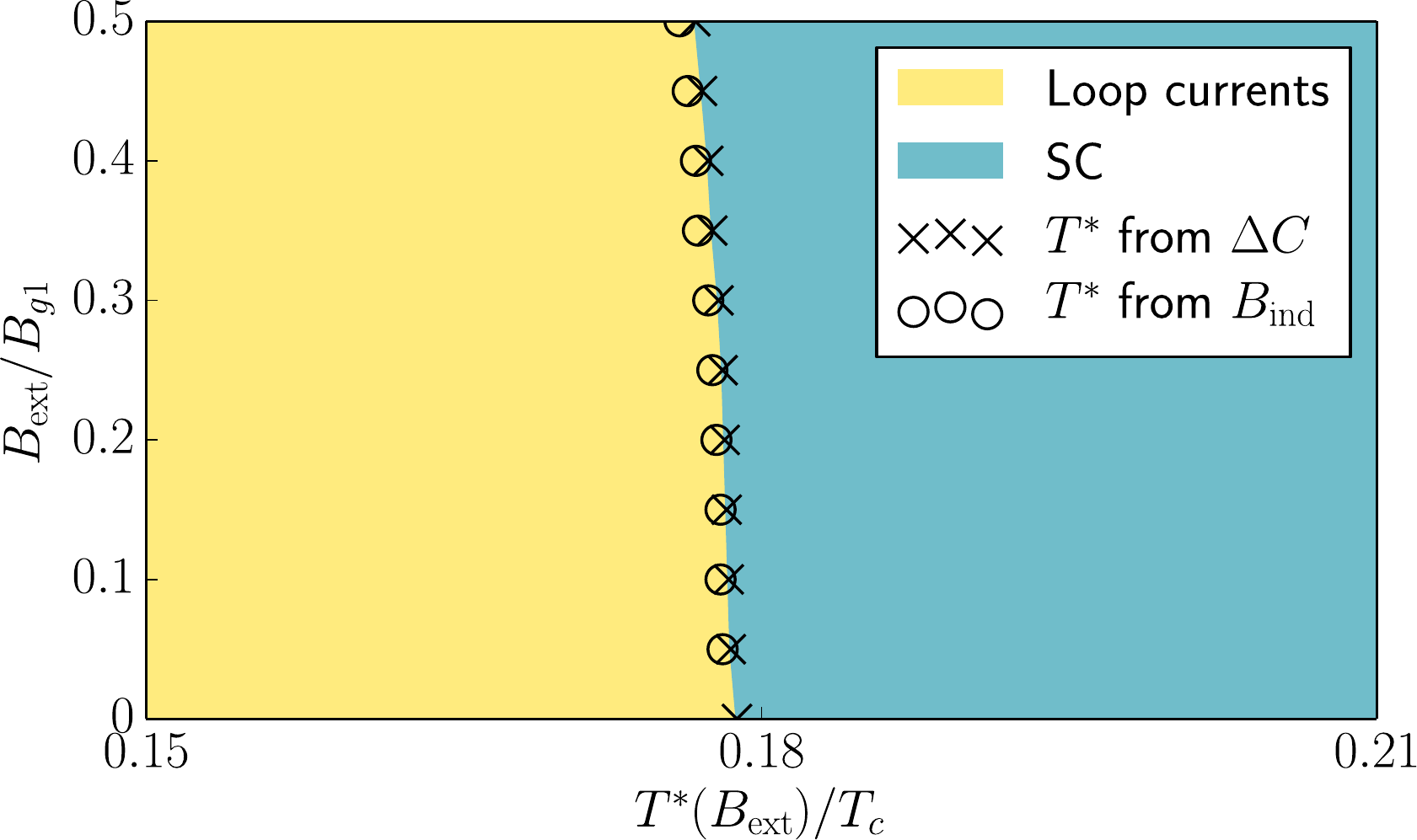}
\caption{\label{fig:phase_diagram} (Color online) Phase diagram of the $d$-wave superconductor with pair breaking edges, showing the transition temperature $T^{*}$ to a state with spontaneously broken continuous translational symmetry versus the external magnetic flux density. The crosses show $T^*$ extraced from the jump in the specific heat in Fig.~\ref{fig:free_E}(e), while the open circles show $T^*$ extracted from the minimum of the total induced magnetic flux in Fig.~\ref{fig:int_B}(a).}
\end{figure}

The knee in the entropy leads to a jump in the specific heat, as shown in Fig.~\ref{fig:free_E}(e) and Fig.~\ref{fig:free_E}(f). The heat capacity is expressed in units of the heat capacity jump at the normal-superconducting phase transition at $T_c$ for a bulk $d$-wave system
\begin{equation}
\label{eq:hc_d}
\Delta C_d = \frac{2\alpha}{3} {\cal A} k_B^2T_cN_F,
\end{equation}
where
\begin{equation}
\label{eq:alpha}
\alpha = \frac{8\pi^2}{7\zeta(3)},
\end{equation}
with $\zeta$ being the Riemann-zeta function. The jump in heat capacity at the phase transition is an edge-to-area effect, and grows linearly as the sample becomes smaller. The jump is roughly $4.5 \%$ of $\Delta C_d$ for the mesoscopic ${\cal A}=60\times 60\xi_0^2$ grain considered here, and grows as the size of the grain is reduced.  The phase transition temperature $T^{*}$ is extracted as a function of $B_{\text{ext}}$ as the midpoint temperature of the jump in the specific heat. Fig.~\ref{fig:phase_diagram} shows a phase diagram where the $T^{*}$, extracted in this way from the specific heat, is plotted versus external field strength (crosses). We compare this with $T^*$ extracted as the minimum [the ``kink'', see Fig.~\ref{fig:int_B}(a)] in the induced flux. The small lowering of $T^{*}$ with increased $B_{\text{ext}}$ is caused by the competing paramagnetic response.

From the above it is clear that the phase with edge loop-currents shows extreme robustness against an external magnetic field in the whole Meissner region ($B_{\text{ext}} \leq B_{g1}$). The magnitude of the spontaneously formed superfluid momentum $\bm{p}_s$ at the edge grows nonlinearly to be very large for $T<T^*$, fueled by the lowering of the free energy by Doppler shifts of the flat band of Andreev surface states. The corresponding correction to $\bm{p}_s$ due to the process of screening of the external magnetic field, is in comparison small. Thereby, $T^*$ is not dramatically shifted in a magnetic field and the symmetry broken phase below $T^*$ is robust.

\section{\label{sec:conclusion}Summary and conclusions}

We have used the quasiclassical theory of superconductivity to study mesoscopic superconducting grains with pair-breaking edges. Using this method, a phase which spontaneously breaks translational symmetry and $\mathcal{T}$-symmetry was found in our previous study\cite{bib:hakansson_2015} and we have in this paper discussed the magnetic field dependent thermodynamics in detail. We have shown that the phase should be quantified in terms of its order parameter, the vector field $\bm{p}_s(\bm{R})$, which contains edge sources and sinks, as well as saddle point disclinations. At these points $\nabla\times\bm{p}_s\neq 0$. We have studied how an external magnetic field in both the Meissner state ($B_{\text{ext}} < B_{c1}$) and the mixed state ($B_{c1} < B_{\text{ext}} < B_{c2}$) affects this phase, and in particular, how the transition temperature $T^{*}$ into this phase varies with the intensity of the external field. Above $T^{*}$, the external field gives rise to the usual diamagnetic Meissner current in the bulk sample, and a paramagnetic response along pair-breaking edges. The paramagnetic current is carried by quasiparticles (midgap states), typically survives a coherence length into the sample, and gives rise to a tiny Doppler shift of mid-gap states that competes with the loop-current phase. As the temperature approaches $T^{*}$, two types of nodes form where the paramagnetic response is locally suppressed and enhanced. As the temperature is lowered below $T^{*}$, current loops appears at the nodes with opposite circulations in neighboring loops. The loop current strength increases highly non-linearly, suppressing the paramagnetic response. As the strength of the external magnetic field increases, the size of the Doppler shift due to the paramagnetic response grows linearly. Therefore, $T^{*}$ decreases slightly as the magnitude of the external field increases. The influence of the external field, and in particular the sudden disappearance of the paramagnetic response, leads to observables which we argue should be visible in experiment. For example the ``kink'' in the total induced flux at the $T^{*}$. The vortices should be directly observable with recently developed scanning probes \cite{bib:vasyukov_2013,bib:pelliccione_2016}, and the sudden disappearance of the paramagnetic response should to be observable with nano-SQUIDS and possibly in penetration-depth experiments. Furthermore, the large jump in heat capacity at the phase transition should be observable with nanocalorimetry \cite{bib:Diao_2016}.

\begin{acknowledgments}
We thank the Swedish Research Council for financial support. It is a pleasure to thank Mikael H\r{a}kansson, Niclas Wennerdal, and Per Rudquist for valuable discussions.
\end{acknowledgments}


%

\end{document}